\documentclass{aa} 
\usepackage{times}
\usepackage{psfig}
\usepackage{epsfig}

\begin{document} 


\title{The infrared properties of the new outburst star
IRAS~05436$-$0007 in quiescent phase}
\author{P. \'Abrah\'am\inst{1} 
        \and 
        \'A. K\'osp\'al\inst{2}
        \and 
        Sz. Csizmadia\inst{1}
        \and 
        A. Mo\'or\inst{1}
        \and
        M. Kun\inst{1}        
        \and 
        G. Stringfellow\inst{3}}
\offprints{P. \'Abrah\'am, email: abraham@konkoly.hu}
\institute{Konkoly Observatory of the Hungarian Academy of Sciences,
           P.O. Box 67, H-1525 Budapest, Hungary 
      \and Department of Astronomy, E\"otv\"os Lor\'and University,
           P.O. Box 32, H-1518 Budapest, Hungary
      \and Center for Astrophysics \& Space Astronomy,
           University of Colorado at Boulder, 389 UCB, Boulder, CO  80309-0389,
           USA}

\date{Received date; accepted date} 
\authorrunning{P. \'Abrah\'am et al.}  
\titlerunning{The infrared properties of IRAS~05436$-$0007 in quiescent phase}
\abstract{We compiled and investigated the infrared/sub-mm/mm SED of
the new outburst star IRAS~05436$-$0007 in quiescent phase.  The star
is a flat-spectrum source, with an estimated total luminosity of
$L_{\rm bol}\,{\approx}\,5.6\,L_{\sun}$, typical of low-mass T\,Tauri stars.  
The derived circumstellar mass of $0.5$\,$M_{\sun}$ is rather high among
low-mass YSOs.  The observed SED differs from the SEDs of typical
T\,Tauri stars and of 4 well-known EXors, and resembles more the SEDs
of FU\,Orionis objects indicating the presence of a circumstellar 
envelope. IRAS~05436$-$0007 seems to be a Class II
source with an age of approximately 4${\times}10^5$\,yr. In this
evolutionary stage an accretion disk is already fully developed,
though a circumstellar envelope may also be present.  Observations of
the present outburst will provide additional knowledge on the source.
 } \maketitle


\section{Introduction} 
\label{sect:Intro} 

On 23 Jan 2004 the amateur astronomer J.W.~McNeil discovered a new
nebula towards the Orion\,B molecular cloud, close to the diffuse
nebulosity Messier 78 (McNeil et al.~2004). 
The object was not visible in either of the two Palomar Surveys
(1951, 1990), but a photograph taken in 1966 for the book 
``The Messier Album''\footnote{http://www.seds.org/messier/} (Mallas \& Kreimer 
1978) shows a bright nebulosity very similar to the one of today. 
Also in the very deep [SII] image of Eisl\"offel and Mundt (1997),
taken in October 1995, parts of the nebula are clearly visible though 
fainter than in 1966. 
The alternation of active and quiescent periods, suggested by these 
earlier observations, indicates that the event, probably the eruption
of a pre-main sequence star, may be similar to the well-known EXor-type 
outbursts.

At infrared and sub-millimetre wavelengths, however, the source was
observable also during the quiescent periods (IRAS, 2MASS, Lis et al.~1999,
Mitchell et al.~2001).  At these wavelengths the emission is due to
thermal radiation of circumstellar dust. The infrared/sub-mm/mm data
offer a possibility to study the circumstellar matter -- which is
likely responsible for the explosion via a sudden rise of the
accretion onto the star (Hartmann \& Kenyon~1996) -- prior to an
outburst.  

In this paper we collect all infrared/sub-mm observations available in
the literature and compile a spectral energy distribution (SED)
representative of the quiescent phase. The SED will be analysed, and
compared with SEDs of pre-main sequence stars, including several known
FUORs and EXors.

\section{Infrared/sub-mm/mm data}
\label{sec:data}

\begin{table}
\begin{center}
\begin{tabular}{cccl}
\hline
$\lambda$  & Flux     &  Obs.    & {\bf Name}, reference                        \\
$\mu$m     & [mJy]    &  date    &                                             \\ 
\hline
1.25       & 1.997    &  1998    & {\bf 2MASS J05461313$-$0006048}                \\
1.65       & 13.94    &  1998    & {\bf 2MASS J05461313$-$0006048}                \\ 
2.17       & 50.96    &  1998    & {\bf 2MASS J05461313$-$0006048}                \\ 
\hline
6.7        & 267      &  1997    & ISOCAM map, ISO\_id: 68901103                \\
14.3       & 559      &  1997    & ISOCAM map, ISO\_id: 68901122                \\ 
\hline
12         & 527      &  1983    & {\bf IRAS 05436$-$0007}, PSC                   \\
25         & 1200     &  1983    & {\bf IRAS 05436$-$0007}, PSC                   \\
60         & 2000     &  1983    & {\bf IRAS 05436$-$0007}, Scanpi                \\ 
\hline
350        & 2500     &  1997    & {\bf LMZ\,12}, Lis et al.~(1999)              \\
850        & 180      &  1998    & {\bf Ori B smm\,55},   Mitchell et al.~(2001) \\
1300       & 93       &  1998    & {\bf LMZ\,12}, Lis et al.~(1999)              \\ 
\hline
\end{tabular}
\caption{Infrared/sub-mm/mm observations of IRAS~05436$-$0007
collected from the literature (Sect.\,\ref{sec:data}). No reddening
correction was applied. We adopt the coordinates of the 2MASS source,
$\alpha_{2000}=5^{\rm h}46^{\rm m}13\fs13$ $\delta_{2000}=-0^{\circ}6'4\farcs8$, as
the position of the outburst star.}
\vspace{-5mm}
\label{tab:sum}
\end{center}
\end{table}

From the literature one can collect several infrared and sub-mm/mm flux
values for the new outburst star:

The 2MASS All-Sky Catalog of Point Sources (Cutri et al.~2003)
contains a source coinciding with the optical position of the new
star. We adopt the coordinates of this source, J05461313$-$0006048, as
the position of the outburst object: $\alpha_{2000}=5^{\rm h}46^{\rm m}13\fs13$,
$\delta_{2000}=-0^{\circ}6'4\farcs8$. Table\,\ref{tab:sum} presents the
2MASS JHK$_{\rm S}$ fluxes, which are all flagged as high quality data in
the catalogue.

In the ISO Data Archive\footnote{http://www.iso.vilspa.esa.es/IDA}
there are two mid-infrared maps, taken by the ISOCAM instrument, which
covers the position of the star. Checking the source lists available
in the FITS file headers, it was possible to identify a point source
coinciding with the 2MASS source.  Its flux densities (produced by
ISO's automatic off-line processing software) are also given in
Tab.\,\ref{tab:sum}.

The outburst star was identified with the IRAS source 05436$-$0007 by
Eisl\"offel \& Mundt (1997). We adopted the 12 and 25$\,\mu$m flux
densities from the IRAS Point Source Catalogue (Tab.\,\ref{tab:sum}).
At longer wavelengths the PSC gives only upper limit, therefore we
determined new IRAS fluxes using the SCANPI Processing Tool at
IPAC\footnote{http://irsa.ipac.caltech.edu/applications/Scanpi}. At
60$\,\mu$m it was possible to extract an estimated $2\pm 1$ Jy, but at
100$\,\mu$m no meaningful flux or upper limit could be determined due
to the complex background of the Orion B molecular cloud. In the
remaining of this paper we use the IRAS name when referring to the new
outburst star.

Three sub-mm/mm measurements can be assigned to IRAS~05436$-$0007 at
350 and 1300$\,\mu$m (Lis et al.~1999) and at 850$\,\mu$m (Mitchell el
al.~2001), taking into account the spatial resolutions of the used
instruments.  At 1300$\,\mu$m the source was unresolved, while at
850$\,\mu$m a ${\sim}20''$ object was visible. The total integrated
fluxes, taken from the papers, as well as the names the authors gave
to the object are given in Tab.\,\ref{tab:sum}.

\section{Quiescent periods of the source} 
\label{sect:period}

IRAS~05436$-$0007 was not visible in either of the two Palomar Surveys
(1951, 1990), showing that the source was in quiescence. In order to
define better the active and quiescent periods of the star, we checked
the photographic archive of the Konkoly Observatory (containing
more than 13000 Schmidt plates) and discovered two -- a blue sensitive and
an approximately visual -- photographs for the area of the outburst
star. From a visual inspection of the plates we concluded that the
source was not visible either in 1964 October or in 1976 October
(the estimated detection limit was $17.5\pm0.5$ magnitude in both 
plates). Thus the eruption recorded in 1966 on the photograph in 
``The Messier Album'' (Sect.~\ref{sect:Intro}) was not long-lasting. 
From the listed data one may probably
conclude that IRAS~05436$-$0007 was mainly in quiescent phase
throughout the second half of the last century.

\section{The infrared SED} 
\label{sect:IRSED}

\begin{figure}
\begin{center} 
\psfig{figure=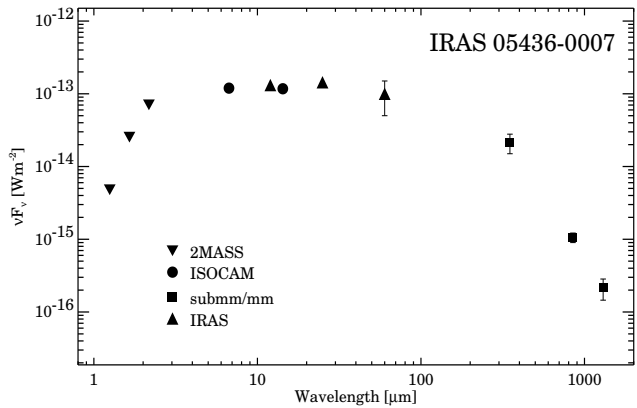,width=55mm,height=85mm, angle=90}
\caption{Spectral energy distributions of IRAS~05436$-$0007. The fluxes are
taken from the literature and are listed in Tab.\,\ref{tab:sum}.}
\label{fig:main}
\end{center} 
\end{figure}

In Fig.\,\ref{fig:main} we plotted all data points listed in
Tab.\,\ref{tab:sum}.  It is not obvious -- due to their different
epochs -- if they form a physically consistent SED representative of
the quiescent phase.  In the near-infrared regime, however, the 2MASS
images do not reveal any extended nebulosity suggesting that the
source was inactive at that date.  In the mid-infrared there exist
observations of two different epochs (IRAS, ISOCAM), whose good
agreement suggests a similar activity level -- probably quiescence --
at the two epochs. At far-infrared wavelengths our study of 7
FU\,Orionis-type stars (K\'osp\'al et al.~2004) suggests that the flux
densities of eruptive stars are practically independent of the
outburst stage, and probably the same is true for the sub-mm/mm
regime.  Thus combining our multiepoch data into a single SED seems to
be justified.  In the following we discuss the different wavelength
regimes of this SED separately.

\begin{figure}
\begin{center} 
\psfig{figure=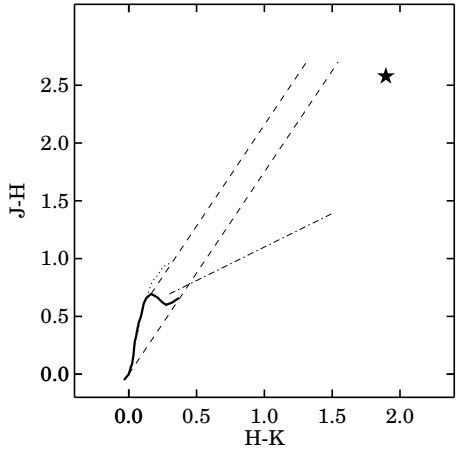,width=60mm,height=60mm, angle=0}
\caption{The location of IRAS~05436$-$0007 ({\it star}) on a
near-infrared colour-colour diagram. Overplotted are the unreddened
main sequence ({\it solid line}) and giant sequence ({\it dotted
line}), the reddening path of the main sequence ({\it dashed line}),
and also the locus of the dereddened classical T Tauri stars ({\it
dashed-dot line}, Meyer et al.~1997).}
\label{fig:JHK}
\end{center} 
\end{figure}

\begin{figure}
\begin{center} 
\psfig{figure=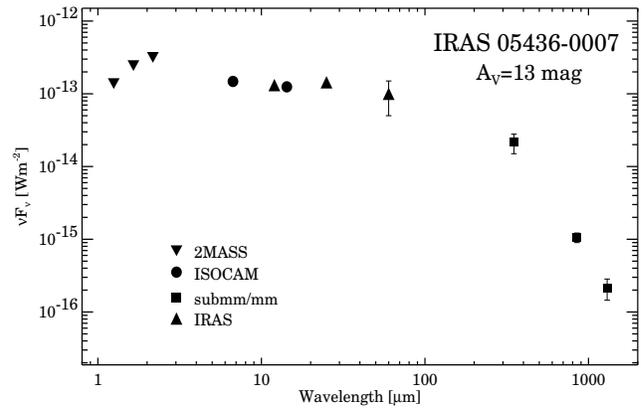,width=55mm,height=85mm, angle=90}
\caption{Spectral energy distributions of IRAS~05436$-$0007 after correcting for
an interstellar extinction of $A_{\rm V}\,{=}\,13\,$magnitude.}
\label{fig:red}
\end{center} 
\end{figure}

\paragraph{Near-infrared:} the unusually high $J{-}H$ and $H{-}K$ indices
indicate that the central source is heavily
extincted. Figure\,\ref{fig:JHK} shows the location of the star on a
$J{-}H$ vs.~$H{-}K$ diagram. A backward projection parallel to the
reddening path onto the locus of the classical T Tauri stars (Meyer et
al.~1997) gives dereddened colour indices of $(J{-}H)_{\rm 0}\,{=}\,1.15$
magnitude and $(H{-}K)_{\rm 0}\,{=}\,1.08$ magnitude. The near-infrared 
reddening is $E(J{-}H)\,{=}\,1.43$ mag, and using the relationship of Rieke \&
Lebofsky (1985) this result gives a visual interstellar extinction of
$A_{\rm V}{\approx}13$ magnitude. 
One should keep in mind that 
the observed near-infrared colours could be affected by scattering within 
the circumstellar environment, therefore the derived extinction value
(and also the result of the reddening correction in Fig.~\ref{fig:red}) should
be taken with some caution.
The $(J{-}H)_{\rm 0}$ and $(H{-}K)_{\rm 0}$ values
indicate the presence of a significant amount of circumstellar matter.

\paragraph{Mid-infrared ($7\,{-}\,25\,\mu$m):} Based on the IRAS and sub-mm data points, 
Lis et al. (1999, their Fig.\,6) modelled the SED as a sum of two
modified Planck functions with $T\,{=}\,165\,$K and $T\,{=}\,30\,$K. The
inclusion of the ISOCAM fluxes, however, does not support this simple
model, but indicates that IRAS~05436$-$0007 is a {\it flat-spectrum
source} at infrared wavelengths. Such a SED requires a less steep
radial temperature profile than expected from an optically thick,
geometrically thin circumstellar disk, and can be the result either of
a flared disk (Adams et al.~1988) or of an extended envelope. Detailed
modelling of FU\,Ori-type objects (Turner et al. 1997) showed that
flared disks alone may not be able to produce a flat spectrum over an
extended wavelength range, thus in many cases envelopes are invoked to
explain the SEDs of the flat-spectrum sources in the infrared.

\paragraph{Far-infrared:} 
the flat spectrum declines at longer wavelengths. This part of the SED
was fitted by Lis et al.~(1999) assuming a modified Planck-function
$B_{\nu}(T){\nu}^{\beta}$ with $T\,{\sim}\,30\,$K and
${\beta}\,{\sim}\,1.0$. Including the 850$\,\mu$m data point of Mitchell
et al.~(2001) does not change the fit.  The turning point in the
spectral shape, which marks the transition between optically thick and
thin emission, is probably between 100 and 200$\,\mu$m.  Since in many
classical T\,Tauri stars the turning point is in the range of 60 to
100$\,\mu$m (e.g. D'Alessio et al.~1999), the fit results indicate the
presence of a large amount of cold material in the outer part of the
IRAS~05436$-$0007 system.  The submillimetre spectral slope,
characterised by $\beta\,{\sim}1$, shows that the average particle
size is larger than the typical value in the interstellar medium
($\beta\,{\sim}2$), probably due to grain evolution in the
circumstellar environment.

In order to compute the luminosity of IRAS~05436$-$0007, we applied a
reddening correction on the SED of Fig.\,\ref{fig:main} assuming
$A_{V}\,{=}\,13$\,mag.  The result is shown in Fig.\,\ref{fig:red}.
Integrating over the corrected SED between 1 and 1300$\,\mu$m gives a
total infrared-to-submillimetre luminosity of $5.6\,L_{\sun}$ (for the
distance of the Ori\,B cloud $d\,{=}\,460\,$pc was adopted).  This
value is higher than the estimate of Lis et al.~(1999,
$L_{\rm bol}\,{\sim}\,2.7\,L_{\sun}$) but the discrepancy is explained by the
inclusion of the new near-infrared data points and by the reddening
correction. The obtained luminosity of $5.6\,L_{\sun}$ clearly shows
that IRAS~05436$-$0007 is a low-mass T Tauri-like object.

From the sub-mm flux densities it is possible to estimate the circumstellar
mass.  Lis et al.~(1999) computed 0.4$\,M_{\sun}$, while from the measurement
of Mitchell et al.~(2001, Eq.~4) we derived ${\sim}\,0.6\,M_{\sun}$. The
obtained mass is rather high compared to typical T\,Tau values of
${\sim}\,0.01\,M_{\sun}$ (e.g. Beckwith et al.~1990).

\section{Discussion}


\subsection{Comparison with pre-main sequence stars}

In this section we compare the SED of IRAS~05436$-$0007 with a sample
of SEDs of pre-main sequence stars. We focus on the
$\lambda\,{\ge}\,10\,\mu$m spectral range, because at
optical/near-infrared wavelengths the variation of circumstellar
extinction with inclination angle introduces a diversity in the
spectral shapes.

\paragraph{T\,Tauri stars.}
The median SED of 39 young stars from the Taurus-Auriga star-forming
region was constructed by D'Alessio et al.~(1999).  Their Fig.~5
reveals that the median SED declines towards longer wavelengths in the
10$-$100$\,\mu$m range, which is clearly inconsistent with the flat
spectrum of IRAS~05436$-$0007.

\paragraph{EXors.}
The infrared SEDs of 4 well-known EXor-type objects (EX\,Lup, DR\,Tau,
UZ\,Tau, VY\,Tau) in quiescent phase were derived from ISOPHOT
observations by Stringfellow et al.~(2004).  Similarly to the T\,Tauri
stars, these objects also exhibit SEDs declining towards longer
wavelengths, significantly differing from the SED of
IRAS~05436$-$0007.
On the other hand, the intermediate-mass young star PV\,Cephei, 
which is sometimes classified as an eruptive EXor-like variable 
(e.g. Teodorani et al. 1999) shows an approximately flat SED (\'Abrah\'am et
al. 2000).

\begin{figure}  
\begin{center} 
\psfig{figure=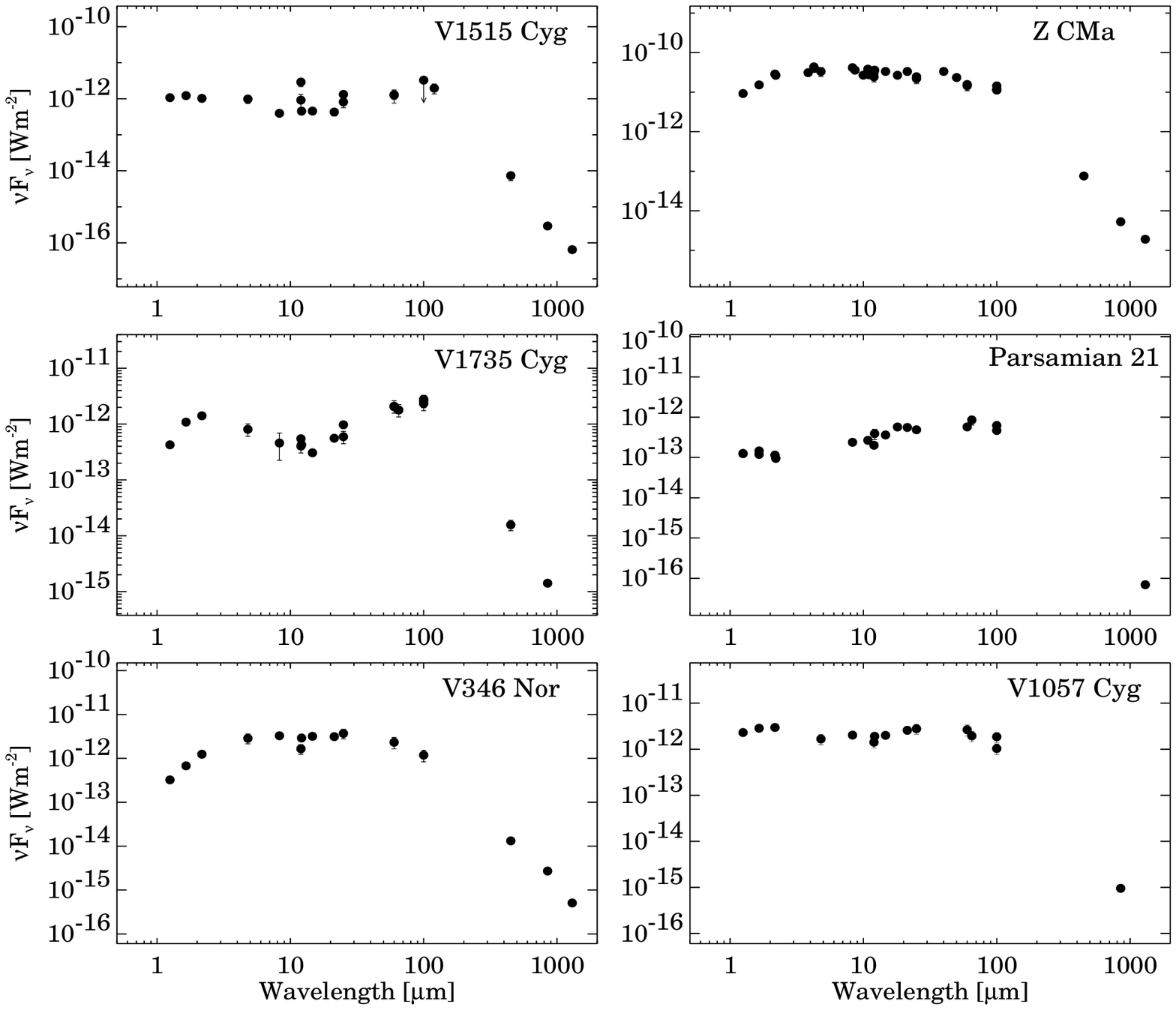,width=88mm,height=76mm}
\caption{Spectral energy distributions of 6 FUORs (from K\'osp\'al et
al.~2004). The presented data are based on
observations obtained by ISOPHOT, the photometer on-board the {\it Infrared
Space Observatory}, supplemented by IRAS, MSX, 2MASS, and sub-mm/mm
data. In case the SED changed between 1983 and ~1997 (K\'osp\'al et
al.~2004) only the more recent data set was plotted.}
\label{fig:fuors}
\vspace{-5mm}
\end{center} 
\end{figure}

\paragraph{FUORs.}
In Fig.\,\ref{fig:fuors} we collected the infrared SEDs of 6
FU\,Ori-type objects published in K\'osp\'al et al.~(2004), which are
in the post-outburst phase. 
All these SEDs are flat (${\nu}F_{\nu}{\sim}{\rm const.}$) or even raising with
increasing wavelength in the
10--100$\mu$m spectral range. These SEDs resemble more the SED of the 
new outburst star (Fig.\,\ref{fig:main}) than do either the SEDs of the  
T\,Tau-stars or the EXors.  
Especially the 3 sources with
flat mid-infrared spectra: V346\,Nor, V1057\,Cyg, and Z\,CMa, look similar 
to IRAS~05436$-$0007 (though in the case of Z\,CMa the companion, a Herbig 
Ae/Be star, might also contribute to the SED).

\paragraph{}
The fact that IRAS~05436$-$0007 was bright also in 1966
(Sect.\,\ref{sect:period}) suggests multiple active periods,
i.e. an EXor-like nature. 
Its luminosity of $5.6\,L_{\sun}$ 
(Sect.\,\ref{sect:IRSED}) is also more typical for the T\,Tau-like EXors than 
for the 5--100 times more luminuous FUORs (Sandell \& Weintraub 2001). 
However, on the basis of the shape of the spectral energy distribution,
IRAS~05436$-$0007 is more similar to the FU\,Orionis objects.
The flat spectra of the FUORs are usually interpreted in terms of
extended circumstellar envelopes (Kenyon \& Hartmann 1991, 
Turner et al. 1997), whose material is falling onto the outer parts of the
accretion disk. Thus IRAS~05436$-$0007 might also be surrounded by an
envelope, which -- together with the unusually large circumstellar mass
of $\approx$0.5$\,M_{\sun}$ 
(Sect.\,\ref{sect:IRSED}) -- would make this object relatively unique among 
the known EXors. Observations of the present outburst, including measurements 
of the infrared SED with the Spitzer Space Telescope, will provide more data
to compute detailed models of the circumstellar structure.


\subsection{Evolutionary stage of IRAS~05436$-$0007}

Based on the submillimetre-to-bolometric luminosity ratio Lis et
al.~(1999) proposed that IRAS~05436$-$0007 is a relatively young and
embedded Class 0 source, though some observations (e.g. the lack of
molecular outflow) seemed to indicate that the source was more
evolved. In order to estimate the evolutionary stage of the source, we
followed the method of Chen et al.~(1995) and computed the bolometric
temperature $T_{\rm bol}$ according to their Eq.\,1.  The resulting
$T_{\rm bol}\,{=}\,830\,$K and the bolometric luminosity
$L_{\rm bol}\,{\approx}\,5.6\,L_{\sun}$ was then compared with the
distribution of corresponding values among YSOs in the Taurus and
$\rho$\,Oph star forming regions (Chen et al.~1995). From this check
we can conclude that IRAS~05436$-$0007 seems to be a Class II object
(close to the Class I/Class II boundary), and its age -- according to
Fig.~4 of Chen et al.~(1995) -- is approximately 4${\times}10^5$\,yr.
\section{Conclusions}

We compiled and investigated the infrared/sub-mm/mm SED of the new
outburst star IRAS~05436$-$0007 in quiescent phase.  The star is a
flat-spectrum source, with an estimated total luminosity of
$L_{\rm bol}\,{\approx}\,5.6\,L_{\sun}$, typical of low-mass T\,Tauri stars.  
The derived circumstellar mass of $0.5\,M_{\sun}$ is rather high among
low-mass YSOs.  The observed SED differs from the SEDs of typical
T\,Tauri stars and of 4 well-known EXors, and resembles more the SEDs
of FU\,Orionis objects indicating the presence of a circumstellar envelope. 
IRAS~05436$-$0007 seems to be a Class II
source with an age of approximately 4${\times}10^5$\,yr. In this
evolutionary stage an accretion disk is already fully developed,
though a circumstellar envelope may also be present.  Observations of
the present outburst will provide additional knowledge on the source.



\end{document}